\documentclass[aps,prl,preprint,superscriptaddress,showpacs]{revtex4}

\usepackage{amsmath} %need this one for the modified franck-condon equation
\usepackage{amsfonts} %For the Imaginary and Real symbols
\usepackage{graphicx}
\usepackage[english]{babel}
\usepackage{epstopdf} %you need this package to submit postscript files electronically to prl and still be able to use pdf LaTex
\begin{document}

\preprint{Phys.\ Rev.\ Lett. \textbf{98}, 206406 (2007) }

% Title of paper -------------
 \title{The role of intermolecular coupling in the photophysics of disordered organic semiconductors: Aggregate emission in regioregular polythiophene}
 
 % Author, affiliation (repeat \author .. \affiliation etc. as needed
 \author{Jenny Clark}
 \affiliation{Cavendish Laboratory, University of Cambridge, J.J.\ Thomson Avenue, Cambridge, CB3~0HE, United Kingdom}
 \author{Carlos Silva}
 \affiliation{D\'{e}partement de physique et Regroupement qu\'{e}b\'{e}cois sur les mat\'{e}riaux de pointe, Universit\'{e} de Montr\'{e}al, C.P.\ 6128, Succursale centre-ville, Montr\'{e}al (Qu\'ebec) H3C~3J7, Canada}
 \author{Richard H.\ Friend}
 \affiliation{Cavendish Laboratory, University of Cambridge, J.J.\ Thomson Avenue, Cambridge, CB3~0HE, United Kingdom}
 \author{Frank C.\ Spano}
 \affiliation{Department of Chemistry, Temple University, Philadelphia, PA 19122}
 %-------the date----------
 \date{\today}
 
 %-------abstract---------
 \begin{abstract}
We address the role of excitonic coulping on the nature of photoexcitations in the conjugated polymer regioregular poly(3-hexylthiophene). By means of temperature-dependent absorption and photoluminescence spectroscopy, we show that optical emission is overwhelmingly dominated by weakly coupled H-aggregates. The relative absorbance of the 0--0 and 0--1 vibronic peaks provides a powerfully simple means to extract the magnitude of the intermolecular coupling energy, approximately 5 and 30\,meV for films spun from isodurene and chloroform solutions respectively.
\end{abstract}
 

\maketitle

%-------------------------------- Introduction -------------------------------------------------

%\section{INTRODUCTION}

Regioregular poly(3-hexylthiophene) (rrP3HT) is a model material to explore the role of supramolecular order in the physics of polymeric semiconductors. In thin films it  self-organizes into two-dimensional $\pi$-stacked lamellar structures~\cite{McCullough_1993, Sirringhaus_1999}. These structures allow for two-dimensional charge-transport, resulting in field-effect mobilities of order 0.1\,cm$^2$(Vs)$^{-1}$~\cite{Sirringhaus_1998}, approaching those of amorphous silicon.

The degree to which rrP3HT forms these supramolecular structures depends upon processing conditions and molecular weight~\cite{Kline_2006_Rev}. Morphology, in turn, affects electronic structure. Photoinduced and charge-modulation spectroscopy have provided evidence for delocalized polarons~\cite{Osterbacka_2000, Brown_2001} but the nature of the primary \emph{neutral} excitation has been the subject of debate~\cite{Sakurai_1997, Brown_2003, Kobayashi_2003, Jiang_2002, Koren_2003, Spano_2005_P3HT}. A reasonable starting point is to consider a simple intrachain exciton~\cite{Sakurai_1997}. However, the absorption and emission spectra cannot be fit using a simple Franck-Condon progression. Brown et al.\ proposed two emissive states close in energy, an \emph{intra}- and an \emph{inter}chain state, both contributing to the absorption and emission spectra~\cite{Brown_2003}. On the other hand, Kobayashi et al.\ proposed two \emph{intra}chain emissive states, one with allowed transition to the ground state ($1B_u$) and one generally forbidden but weakly allowed due to vibronic coupling ($2A_g$)~\cite{Kobayashi_2003}. Others proposed that the first excited state is a single \emph{inter}chain H-aggregate-type state~\cite{Jiang_2002, Koren_2003}. One of us has recently built upon the latter idea with a theoretical model of \emph{weakly interacting} H-aggregate states~\cite{Spano_2005_P3HT}. We conclude here that this model describes comprehensively the photophysics of rrP3HT.  

We are able to describe the rrP3HT absorption and emission spectra by invoking a single emitter and without the need to invoke more complicated models~\cite{Brown_2003, Kobayashi_2003}. This is of general interest in other polymeric semiconductor systems exhibiting apparent non-Condon emission~\cite{Ho_2001, Meskers_2000_CP,Rothberg02,Schwartz03}. Using an analytical form of a weakly-interacting H-aggregate model, we describe unambiguously the shape of the spectra from a micro-morphological point of view. We demonstrate that the room-temperature bulk absorption spectrum contains microscopic detail of intramolecular order and interchain coupling energy in the solid state.

%------------------------- Experimental ---------------------------------
Solutions of rrP3HT (Plextronics) were made in isodurene and anhydrous chloroform (Aldrich) and heated to 70$^{\circ}$C for 30 min.\ in N$_2$ atmosphere. Films were spin-coated in N$_2$ atmosphere from hot solution and were subsequently heated to 100$^{\circ}$C for 30 minutes. Absorption spectra were measured with a Hewlett-Packard UV-VIS spectrometer in ambient conditions. Time-correlated single-photon counting was used to measure the photoluminescence (PL) lifetime of all samples, as well as the time-integrated  PL spectra of the solutions using an apparatus described in detail elsewhere~\cite{Morteani_2003} and a pulsed diode laser (407\,nm; 70\,ps FWHM) as the excitation source. An argon-ion laser (488\,nm) was used to measure steady-state PL spectra in films in a continuous-flow cryostat using an Oriel IV Instaspec spectrograph. Films of different thickness were measured to rule out self-absorption effects. Photoluminescence excitation (PLE) spectra were measured in vacuum with a Varian Cary Eclipse Fluorescence Spectrophotometer.
%-----------------------------------------------------------------------------------------------

\begin{figure}[t]
\includegraphics[width=0.46\textwidth]{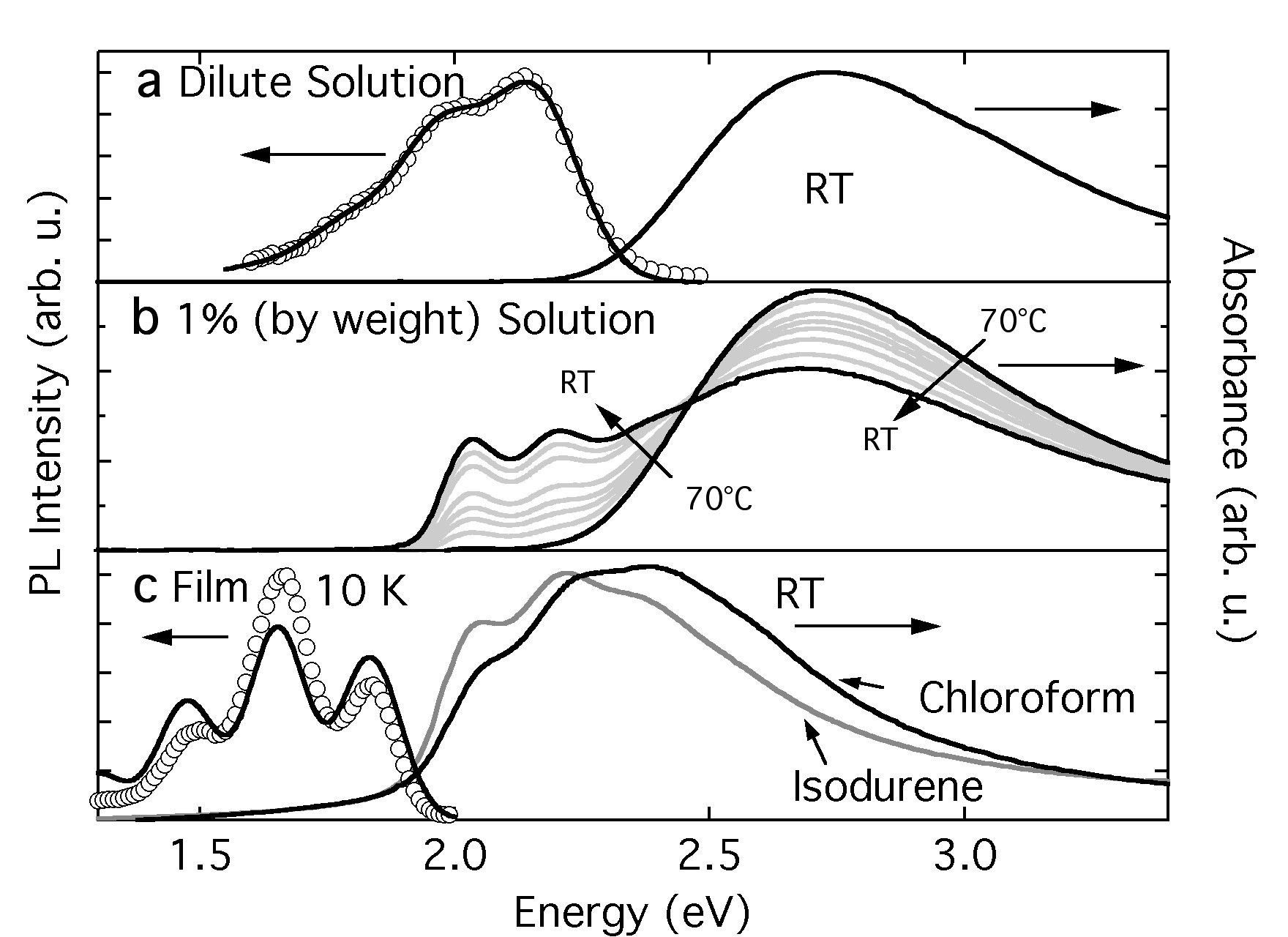} 
	\caption{(a) Normalized room-temperature (RT) absorption and PL spectra of rrP3HT in a 0.0001\% wt isodurene solution. Solid line through PL is a Franck-Condon fit. (b) Temperature-dependent absorption spectra of 1\% wt isodurene solution. (c) Normalized absorption spectra of a film spun from 1\% wt isodurene and chloroform solutions. The PL spectrum of a chloroform film is also shown (open circles) with a Franck-Condon best-fit (solid curve). The PL spectrum was measured at 10\,K whereas absorption spectra were measured at RT. }
	\label{Fig1}
\end{figure}
Fig.~\ref{Fig1}(a) shows room-temperature (RT) PL and absorption spectra of a 0.0001\% by weight  solution of rrP3HT in isodurene (a `bad' solvent). At such low concentration we do not observe a solvent dependence on the spectral shape. These spectra are attributed to \emph{intra}chain excitations and are therefore reproduced with a Franck-Condon model where the relative intensity of the vibronic replica is given by 
	\begin{equation}
		I_{0\to m} \propto (\hbar \omega)^3 n_f^3 \frac{S^{m} \exp{(-S)}}{m!}
		\label{Frank_Condon_Princinple}
	\end{equation}
where $n_f$ is the real part of the refractive index at photon energy $\hbar \omega$, $m$ denotes the vibrational level. $S$ is the Huang-Rhys factor, which gives a measure of the coupling between the electronic transition and a phonon mode~\cite{Ho_2001}. 

We fit the dilute PL spectrum in Fig.~\ref{Fig1}(a) assuming that the C=C symmetric stretch at 0.18\,eV dominates the coupling to the electronic transition~\cite{Botta_1992,Louarn_1996}. For simplicity we used a Gaussian lineshape, with the same width for each of the vibronic transitions. The refractive index of the solvent was taken to be constant across the examined range~\cite{Samoc_2003}. The fit is shown as the solid line in Fig.~\ref{Fig1}(a) and gives $S=1.00 \pm 0.05$.

At significantly higher concentration (1\% wt) the absorption spectrum exhibits additional, structured components at lower energy, which disappear upon heating the solution to $70^\circ$C (Fig.~\ref{Fig1}(b)). The emergence of the red-shifted species is correlated with the formation of crystallites as the chains extend and planarize~\cite{Rumbles_1996, Yue_1996, Theander_2001}. A corresponding reduction in PL efficiency \cite{Rumbles_1996} suggests formation of weakly emissive aggregates. The low-energy absorption observed in solution at high concentration dominates the spectrum in the solid state (Fig.~\ref{Fig1}(c)). The absorption spectrum of a film spun from chloroform, a good solvent, is similar to that spun from isodurene, but the relative intensity of the 0--0 and 0--1 peaks is solvent-dependent. 
We note that we cannot fit the solid-state PL spectrum using a standard single-oscillator Franck-Condon progression at any temperature studied (in the range 10--300\,K), as shown in Fig~\ref{Fig1}(c) for the spectrum at 10\,K (see also Ref.~\cite{Brown_2003}). We conclude from these observations that most of the polymer chains self-organize to form aggregates in the film.

\begin{figure}[t]
\includegraphics[width=0.5\textwidth]{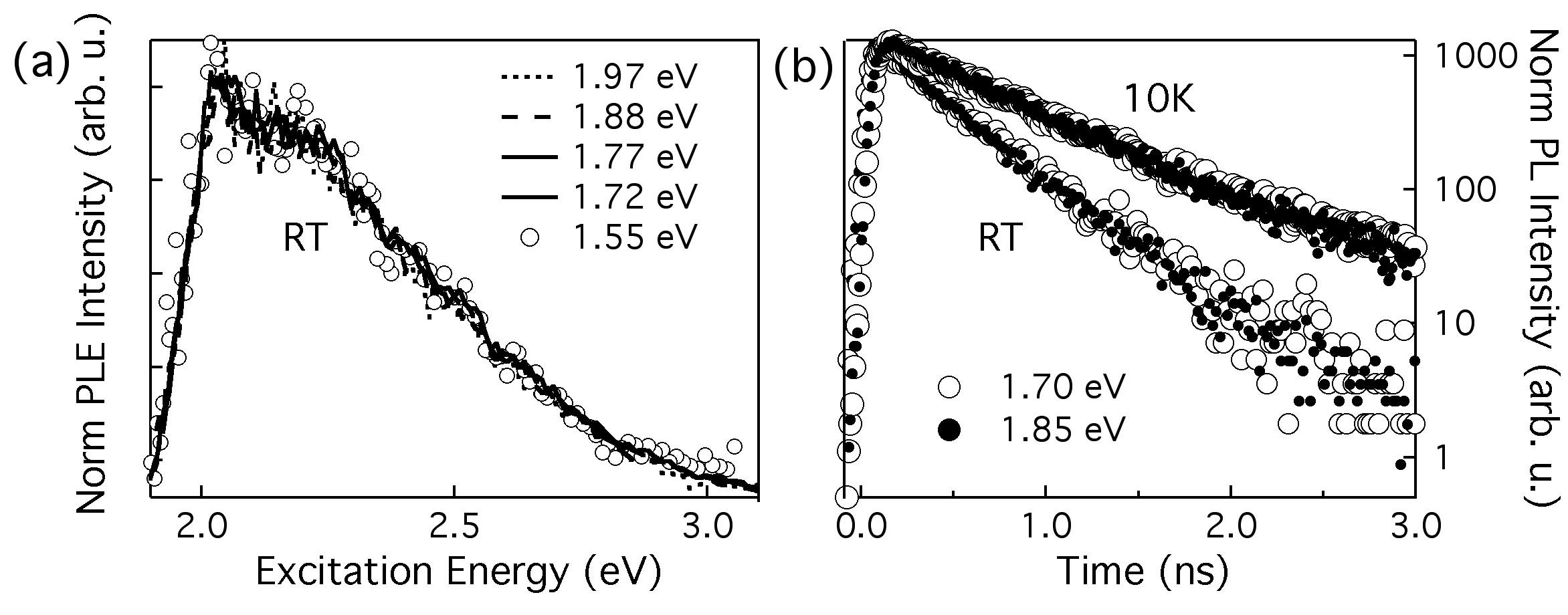}
	\caption{(a) Normalized PLE spectra at room temperature (RT) of a film spun from a chloroform solution at different detection photon energies. (b) PL decay at two photon energies measured at RT and 10\,K.}
	\label{Fig2}
\end{figure}

We now consider whether there is evidence for more than one emissive species. The solid-state PL excitation (PLE) spectra (Fig.~\ref{Fig2}(a)) show no dependence on detection wavelength and resemble the absorption spectrum. The PL lifetime is the same at the 0--0 and 0--1 energies at all temperatures (Fig.~\ref{Fig2}(b)). We conclude that there is a single emissive species in the solid state.

\begin{figure}[t]
\includegraphics[width=0.46\textwidth]{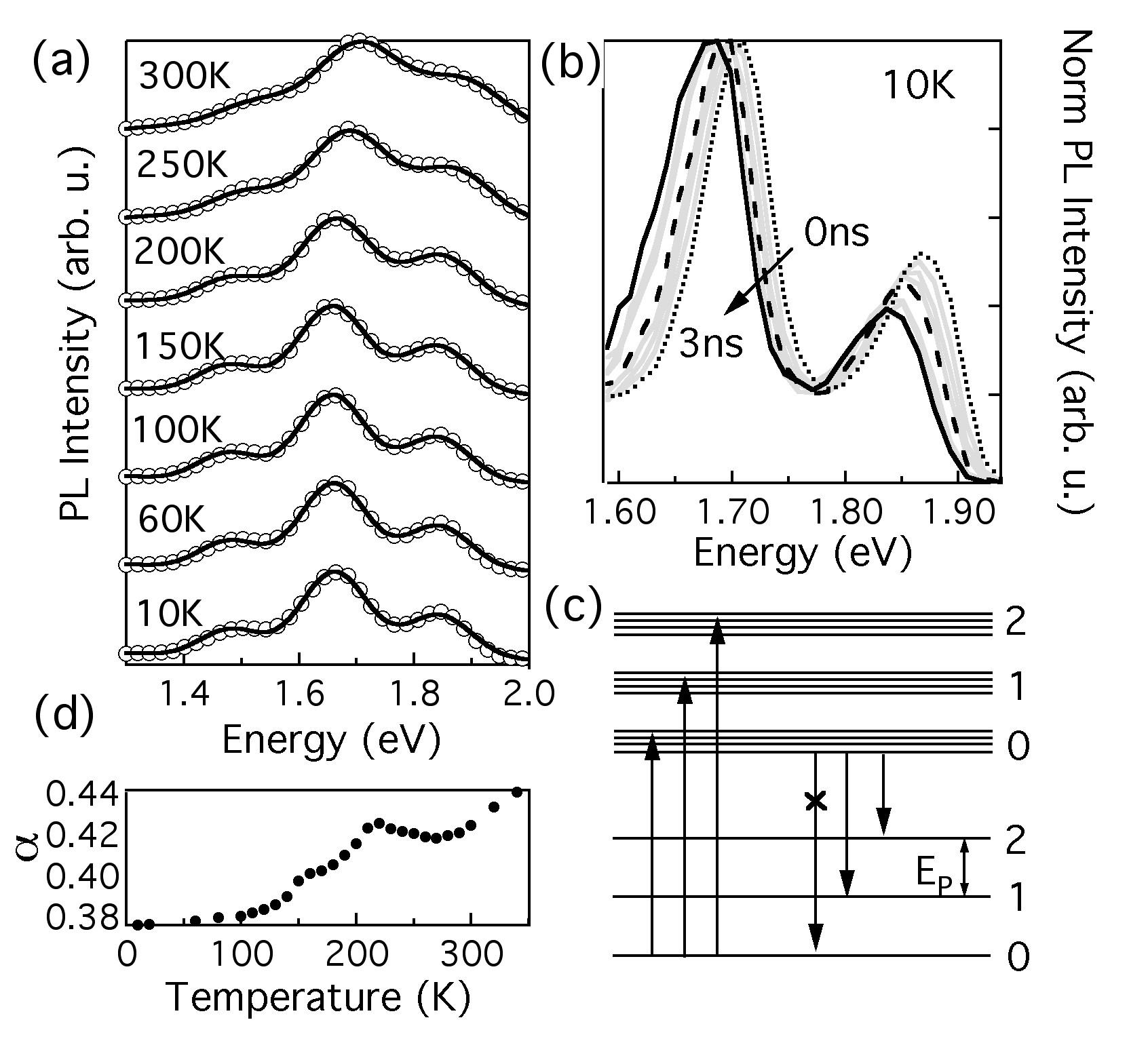}
	\caption{(a) PL spectra of a film spun from chloroform solution at various temperatures. The solid lines through the data are fits to Eq.~\ref{Modified_FC}. (b) Time-resolved PL spectra measured at 10\,K, normalized to the 0--1 peak. Spectra are integrated over 0.25-ns windows and presented here from 0--3\,ns. (c) Jablonski diagram showing the exciton energy level structure.  (d) Variation of $\alpha$ with temperature (see Eq.~\ref{Modified_FC}). 
	}
	\label{Fig3}
\end{figure}

To elucidate the nature of the emissive species, we measured temperature-dependent and time-dependent PL spectra (Fig.~\ref{Fig3}). Part (a) shows temperature-dependent PL spectra of a film from 10\,K to 300\,K. As the temperature increases, the spectra blue-shift and broaden and the 0--0 peak becomes relatively more intense. Temperature-dependent PL spectra of \emph{intramolecular} excitations generally show a blue-shift and broadening upon increasing temperature, but no change in relative intensities of vibronic peaks~\cite{Ho_2001}. Fig.~\ref{Fig3}(b) displays time-dependent PL spectra at 10\,K, normalized to the 0--1 peak intensity. The 0--0 peak becomes relatively less intense with time, and we observe a dynamic spectral red-shift. 

We demonstrate here that such temperature dependence and temporal evolution is consistent with an H-aggregate model in the weak intermolecular coupling regime. 
In H-aggregates the transition from (to) the ground state to (from) the first excited state is forbidden. Intermolecular coupling is taken to be resonant coulomb (excitonic) coupling, denoted $J$. At 3.8\,\AA~$\pi$-stacking distance in rrP3HT \cite{Sirringhaus_1999}, we might expect a large $J$~\cite{Cornil_1998}, however, the long conjugation length in rrP3HT (estimated to be between 20--25 thiophene units from X-ray studies~\cite{Sirringhaus_1999}) also plays an important role. Quantum chemical calculations have shown that increasing the conjugation length reduces $J$~\cite{Manas_1998, Beljonne_2000}. This has been recently demonstrated experimentally in oligothiophenes~\cite{Westenhoff_2006_AdvMater}, 
 leading to the conclusion that in the long polythiophene chain $J$ could be small.
In the `weak' excitonic coupling regime $J$ is smaller than the vibrational relaxation energy and the interchain interaction leads to vibronic bands as shown in Fig.~\ref{Fig3}(c). The free exciton bandwidth is  $W=4J$~\cite{Spano_2005_P3HT}.

In such an H-aggregate, absorption of light is to the top of the vibronic bands. In accordance with Kasha's rule, emission is from the lowest excited state, with zero oscillator strength for the 0--0 transition in the absence of disorder. By contrast, the side-band transitions are allowed. In the weak coupling limit the intensities of the
0-1 and subsequent replicas are somewhat diminished from their single
molecule values~\cite{Spano_2005_P3HT, Meskers_2000_CP}, consistent with the aggregation-induced reduction in quantum yield. However, the side-bands retain the same \emph{relative} intensities as the single molecule spectrum~\cite{Spano_2005_P3HT}. Hence, the aggregate emission spectrum should have a weak, disorder-allowed 0--0 component followed by a Franck-Condon progression of side-bands. Accordingly, the PL spectrum can be modeled as a modified Franck-Condon progression with a variable 0--0 amplitude.
\begin{equation} \label{Modified_FC}
  \begin{split}
 		&I(\omega) \propto  (\hbar\omega)^{3} n_{f}^{3}  e^{-S}  \times
		\\* 
		&\left[\alpha \Gamma(\hbar\omega-E_{0}) + \sum_{m=1}\frac{S^{m}}{m!}\Gamma(\hbar\omega - (E_{0}-mE_{p}))\right]
  \end{split}
	\end{equation}
We define $n_f$, $m$, and $S$ as in Eq.~\ref{Frank_Condon_Princinple}. $E_0$ is the 0--0 transition energy, $E_{p}$ is the phonon energy of the C=C symmetric stretch (0.18\,eV), $\Gamma$ is the line-shape function (simplified to be purely Gaussian with constant width), and $\alpha$ is a constant whose amplitude is allowed to vary during the fit. As shown in~\cite{Spano_2005_P3HT}, $\alpha$ is a strong function of the disorder width and spatial correlation length.
Fits with Eq.~\ref{Modified_FC} are shown as solid curves in Fig.~\ref{Fig3}(a) with $S=1$ (5\% uncertainty) at all temperatures, $\alpha$ as a function of temperature is shown in Fig.~\ref{Fig3}(c). We note that $S\approx1$ in non-interacting, planar polythiophene chains~\cite{Theander_2001} and should not change upon aggregation in the weak coupling regime. Upon increasing temperature the PL spectra blue-shift and broaden due to increasing thermal disorder.  
We also observe a systematic increase of the 0--0 peak intensity ($\alpha$) with temperature due primarily to thermally excited, less-forbidden levels higher up in the lowest vibronic band, a distribution typical of H-aggregates. 
~Thermal disorder may also lead to a relaxation of the selection rule making the 0--0 transition more allowed. Invoking the first mechanism with the model of Ref.~\cite{Spano_2005_P3HT} leads to a spectral blue-shift and 0--0 peak enhancement with temperature in good qualitative agreement with experiment.
Furthermore, the dynamic red-shift (even at 10\,K) and the loss of 0--0 peak intensity with time indicates energy diffusion to more ordered domains which carry lower 0--0 intensity. We conclude that \emph{H-aggregates are the only emissive species in rrP3HT}.

Having established that weak intermolecular electronic coupling gives rise to the spectral properties presented here, we can use these properties to quantify the magnitude of the electronic coupling. Within an H-aggregate model, the magnitude of the interchain coupling can be estimated from the ratio of the 0--0 and 0--1 absorbance peaks~\cite{Spano_2005_P3HT} (Fig.~\ref{Fig1}(c)). With $S=1$, one finds
\begin{equation}
		\frac{A_{0-0}}{A_{0-1}} \approx 
		\frac{n_{0-1}}{n_{0-0}} 
		\left( \frac{1-0.24W/E_{p}}{1+0.073W/E_{p}}\right)^2 
		\label{Spano_Equation}
	\end{equation}
where $n_{0-i}$ is the real part of the refractive index at the 0--$i$ peak. $E_{p}$ is the phonon energy of the main oscillator coupled to the electronic transition. Using $E_p=0.18$\,eV, a refractive index ratio of $\sim$0.97~\cite{Brown_2003}, and the experimental absorbance ratios shown in Fig.~\ref{Fig1}(c), Eq.\ \ref{Spano_Equation} gives the free exciton bandwidths, $W$, of $\sim$120\,meV for the film spun from chloroform and $\sim$20\,meV from isodurene. We note that the absorbance vibronic peak spacing predicted by the model is not uniform owing to second order effects~\cite{Spano_2006_P3HT}. This is in qualitative agreement with experimental data~\cite{Brown_2003}.

Using this analysis, we can construct a microscopic picture of polymer structure. Assuming similar interchain order in these films, the conjugation length and \emph{intra}chain order play a role in the degree of interchain coupling~\cite{Spano_2005_P3HT}. Remembering that increasing the conjugation length reduces  $J$, we see that the chloroform-spun film, with $J\sim$30\,meV,  displays lower intrachain order than the isodurene film, $J\sim$5\,meV.  Furthermore, $J$ can be used to determine the average conjugation length by comparing it to quantum mechanical calculations on pairs of variable-sized oligomers. Crude estimates implementing Zerner's Intermediate Neglect of Differential Overlap (ZINDO) method predict the conjugation length in chloroform films to be 20--30 monomers. Isodurene films, with significantly smaller $J$, are expected to have conjugation lengths at least twice as large.

Field-effect transistors fabricated with rrP3HT films spun from high boiling point solvents, such as isodurene, display increased $A_{0-0}/A_{0-1}$ ratio and higher field-effect mobility than those from lower boiling point solvents such as chloroform~\cite{Chang_2006}. This indicates that intramolecular order, which is related to $J$, is of overriding importance to optimize charge mobility. This simple procedure for estimating $J$, and thus the conjugation length and degree of intrachain order, is a powerful tool for materials development and processing efforts.
 
 \begin{figure}[t]
\includegraphics[width=0.44\textwidth]{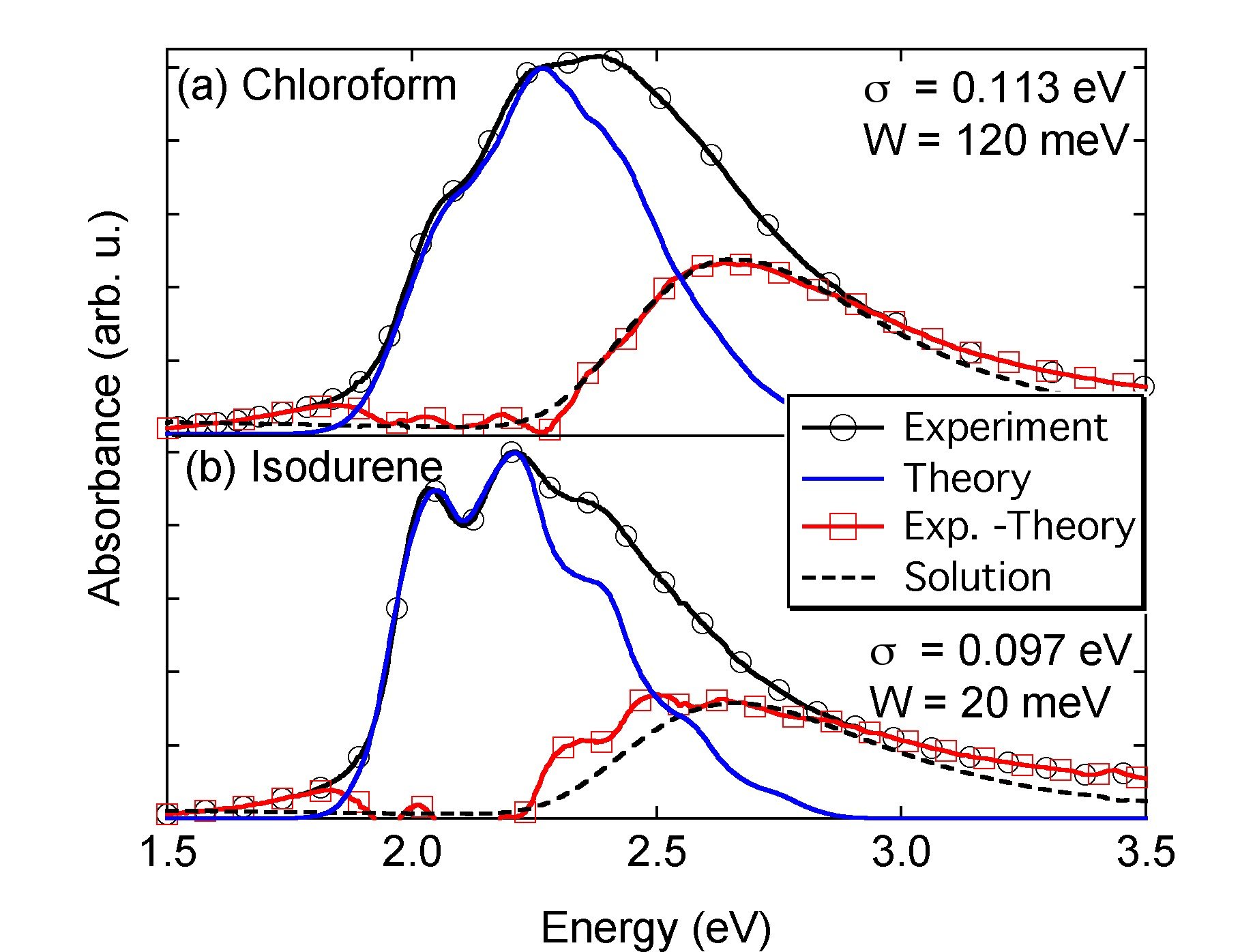}
	\caption{(Color online) Experimental and simulated absorption spectra for films spun from (a) chloroform and (b) isodurene. Calculations were performed on $\pi$-stacks containing 12 polymers, large enough to ensure spectral convergence. Transition frequency offsets were chosen randomly from a Gaussian distribution with (1/e) full width of $\sigma$ and spatial correlation length $l_0$=3. The final absorption spectrum was averaged over 1500 disorder configurations. For further details see~\cite{Spano_2005_P3HT}. }
	\label{Fig4}
\end{figure}

To further explore the physical validity of the weakly interacting H-aggregate model, we present a more detailed analysis of the absorption spectra of films spun from chloroform and isodurene (Fig.~\ref{Fig4}). Alongside the experimental spectra we also plot the calculated spectra using a Holstein-like Hamiltonian with correlated site disorder~\cite{Spano_2005_P3HT}. The solid curves (no markers) are the attenuation, $\mathbb{I}\mathrm{m} \left[\epsilon(\omega)\right]/\mathbb{R}\mathrm{e}\left[n(\omega)\right]$, where the numerator is the imaginary part of the dielectric constant and the denominator is the real part of the refractive index. The evaluation of $\mathbb{I}\mathrm{m} \left[\epsilon(\omega)\right]$ (denoted $A(\omega)$ in~\cite{Spano_2005_P3HT}) is described in detail in Ref.~\cite{Spano_2005_P3HT}. For the frequency dependent index we utilized the measured value. The ratio of the 0--0 to 0--1 peak absorbance is in excellent agreement with the measured ratio, a consequence of the validity of Eq.~\ref{Spano_Equation}. The stronger absorption measured for energies greater than approximately 2.3\,eV is due to unaggregated molecules or short oligomers, both of which are substantially blue shifted compared to the $\pi$-stacked aggregate. To test this idea we subtracted the theoretical curves from the experimental ones, obtaining the differential curves (square markers). These agree very well with the solution spectra in both samples, suggesting an important contribution of unaggregated polymers in the absorption spectra.  However, as demonstrated in this letter, a single species comprised of aggregates is responsible for emission. 

We have shown that the emission in rrP3HT arises from weakly coupled H-aggregates with no significant contribution from intrachain excitons. We can probe the microscopic order in thin films of rrP3HT using the  0--0 and 0--1 peaks of the PL and absorption spectra. This provides an exciting opportunity to explore the interplay between processing, order and ultimately device performance, and a general framework to unravel interchain effects in semiconductor polymers~\cite{Ho_2001,Meskers_2000_CP,Rothberg02,Schwartz03}.

\begin{acknowledgments}
JC was supported by EPSRC and Seiko Epson UK, and thanks D. Russell and A. K\"{o}hler for encouragement and guidance. 
CS is supported by NSERC and the Canada Research Chairs Programme.  
FCS is supported by the NSF, Grant DMR N$^0$~0606028.
\end{acknowledgments}


\begin{thebibliography}{27}
\expandafter\ifx\csname natexlab\endcsname\relax\def\natexlab#1{#1}\fi
\expandafter\ifx\csname bibnamefont\endcsname\relax
  \def\bibnamefont#1{#1}\fi
\expandafter\ifx\csname bibfnamefont\endcsname\relax
  \def\bibfnamefont#1{#1}\fi
\expandafter\ifx\csname citenamefont\endcsname\relax
  \def\citenamefont#1{#1}\fi
\expandafter\ifx\csname url\endcsname\relax
  \def\url#1{\texttt{#1}}\fi
\expandafter\ifx\csname urlprefix\endcsname\relax\def\urlprefix{URL }\fi
\providecommand{\bibinfo}[2]{#2}
\providecommand{\eprint}[2][]{\url{#2}}

\bibitem[{\citenamefont{McCullough et~al.}(1993)\citenamefont{McCulloughet al.}}]{McCullough_1993}
\bibinfo{author}{\bibfnamefont{R.~D.} \bibnamefont{McCullough}}
    \bibnamefont{et al.}, \bibinfo{journal}{J. Am. Chem.
  Soc.} \textbf{\bibinfo{volume}{115}}, \bibinfo{pages}{4910}
  (\bibinfo{year}{1993}).

\bibitem[{\citenamefont{Sirringhaus et~al.}(1999)\citenamefont{Sirringhaus,
  Brown, and Friend}(1999)}]{Sirringhaus_1999}
\bibinfo{author}{\bibfnamefont{H.}~\bibnamefont{Sirringhaus}}~\bibnamefont{et al.},
  \bibinfo{journal}{Nature} \textbf{\bibinfo{volume}{401}},
  \bibinfo{pages}{685} (\bibinfo{year}{1999}).

\bibitem[{\citenamefont{Sirringhaus,
  Tessler, and Friend}(1998)}]{Sirringhaus_1998}
\bibinfo{author}{\bibfnamefont{H.}~\bibnamefont{Sirringhaus}},
\bibinfo{author}{\bibnamefont{N.}~\bibnamefont{Tessler}} \bibnamefont{and}
\bibinfo{author}{\bibnamefont{R.~H.}~\bibnamefont{Friend}},
  \bibinfo{journal}{Science} \textbf{\bibinfo{volume}{280}},
  \bibinfo{pages}{1741} (\bibinfo{year}{1998}).

\bibitem[{\citenamefont{Kline and McGehee}(2006)}]{Kline_2006_Rev}
\bibinfo{author}{\bibfnamefont{R.~J.} \bibnamefont{Kline}}~\bibnamefont{and}
  \bibinfo{author}{\bibfnamefont{M.~D.} \bibnamefont{McGehee}},
  \bibinfo{journal}{Polymer Reviews} \textbf{\bibinfo{volume}{46}},
  \bibinfo{pages}{27} (\bibinfo{year}{2006}).

\bibitem[{\citenamefont{Osterbacka et~al.}(2000)\citenamefont{Osterbacka, An,
  and Jiang}}]{Osterbacka_2000}
\bibinfo{author}{\bibfnamefont{R.}~\bibnamefont{Osterbacka}} \bibnamefont{et al.},
  \bibinfo{journal}{Science} \textbf{\bibinfo{volume}{287}},
  \bibinfo{pages}{839} (\bibinfo{year}{2000}).

\bibitem[{\citenamefont{Brown et~al.}(2001)\citenamefont{Brown, Sirringhaus,
  and Harrison}}]{Brown_2001}
\bibinfo{author}{\bibfnamefont{P.~J.} \bibnamefont{Brown}} \bibnamefont{et al.}, \bibinfo{journal}{Phys. Rev. B}
  \textbf{\bibinfo{volume}{63}}, \bibinfo{pages}{125204}
  (\bibinfo{year}{2001}).

\bibitem[{\citenamefont{Sakurai et~al.}(1997)\citenamefont{Sakurai, Tachibana,
  and Shiga}}]{Sakurai_1997}
\bibinfo{author}{\bibfnamefont{K.}~\bibnamefont{Sakurai}}  \bibnamefont{et al.},
  \bibinfo{journal}{Phys. Rev. B} \textbf{\bibinfo{volume}{56}},
  \bibinfo{pages}{9552} (\bibinfo{year}{1997}).

\bibitem[{\citenamefont{Brown et~al.}(2003)\citenamefont{Brown, Thomas, and
  Kohler}}]{Brown_2003}
\bibinfo{author}{\bibfnamefont{P.~J.} \bibnamefont{Brown}} \bibnamefont{et al.}, \bibinfo{journal}{Phys. Rev. B}
  \textbf{\bibinfo{volume}{67}}, \bibinfo{pages}{064203}
  (\bibinfo{year}{2003}).

\bibitem[{\citenamefont{Kobayashi et~al.}(2003)\citenamefont{Kobayashi,
  Hamazaki, and Kunugita}}]{Kobayashi_2003}
\bibinfo{author}{\bibfnamefont{T.}~\bibnamefont{Kobayashi}} \bibnamefont{et al.},
  \bibinfo{journal}{Phys. Rev. B} \textbf{\bibinfo{volume}{67}},
  \bibinfo{pages}{205214} (\bibinfo{year}{2003}).

\bibitem[{\citenamefont{Jiang et~al.}(2002)\citenamefont{Jiang, Osterbacka, and
  Korovyanko}}]{Jiang_2002}
\bibinfo{author}{\bibfnamefont{X.~M.} \bibnamefont{Jiang}} \bibnamefont{et al}, \bibinfo{journal}{Adv. Funct. Mater.}
  \textbf{\bibinfo{volume}{12}}, \bibinfo{pages}{587} (\bibinfo{year}{2002}).

\bibitem[{\citenamefont{Koren et~al.}(2003)\citenamefont{Koren, Curtis, and
  Francis}}]{Koren_2003}
\bibinfo{author}{\bibfnamefont{A.~B.} \bibnamefont{Koren}}
  \bibnamefont{et al.}, \bibinfo{journal}{J. Am. Chem.
  Soc.} \textbf{\bibinfo{volume}{125}}, \bibinfo{pages}{5040}
  (\bibinfo{year}{2003}).

\bibitem[{\citenamefont{Spano}(2005)}]{Spano_2005_P3HT}
\bibinfo{author}{\bibfnamefont{F.~C.} \bibnamefont{Spano}},
  \bibinfo{journal}{J. Chem. Phys.}
  \textbf{\bibinfo{volume}{122}}, \bibinfo{pages}{234701}
  (\bibinfo{year}{2005}).

\bibitem[{\citenamefont{Ho et~al.}(2001)\citenamefont{Ho, Kim, and
  Tessler}}]{Ho_2001}
\bibinfo{author}{\bibfnamefont{P.~K.~H.} \bibnamefont{Ho}} \bibnamefont{et al.},
  \bibinfo{journal}{J. Chem. Phys.}
  \textbf{\bibinfo{volume}{115}}, \bibinfo{pages}{2709} (\bibinfo{year}{2001}).

\bibitem[{\citenamefont{Meskers et~al.}(2000)\citenamefont{Meskers, Janssen,
  and Haverkort}}]{Meskers_2000_CP}
\bibinfo{author}{\bibfnamefont{S.~C.~J.} \bibnamefont{Meskers}}  \bibnamefont{et al.}, \bibinfo{journal}{Chem. Phys.}
  \textbf{\bibinfo{volume}{260}}, \bibinfo{pages}{415} (\bibinfo{year}{2000}).
  
   \bibitem[{\citenamefont{Rothberg}(2002)\citenamefont{Rothberg}}]{Rothberg02}
\bibinfo{author}{\bibfnamefont{L.~J.} \bibnamefont{Rothberg}},
  \bibinfo{journal}{Proc. Int. School Phys. ``Enrico Fermi", Course CXLIX. Eds. V. M. Agronavich and G. C. La Rocca, Amsterdam}
  (\bibinfo{year}{2002}).
  
   \bibitem[{\citenamefont{Schwartz}(2003)\citenamefont{Schwartz}}]{Schwartz03}
\bibinfo{author}{\bibfnamefont{B.~J.} \bibnamefont{Schwartz}},
  \bibinfo{journal}{Annu. Rev. Phys. Chem.} \textbf{\bibinfo{volume}{54}},
  \bibinfo{pages}{141}
  (\bibinfo{year}{2003}).

\bibitem[{\citenamefont{Morteani et~al.}(2003)\citenamefont{Morteani, Dhoot,
  and Kim}}]{Morteani_2003}
\bibinfo{author}{\bibfnamefont{A.~C.} \bibnamefont{Morteani}} \bibnamefont{et al.},
  \bibinfo{journal}{Adv. Mater.} \textbf{\bibinfo{volume}{15}},
  \bibinfo{pages}{1708} (\bibinfo{year}{2003}).

\bibitem[{\citenamefont{Botta et~al.}(1992)\citenamefont{Botta, Luzzati, and
  Tubino}}]{Botta_1992}
\bibinfo{author}{\bibfnamefont{C.}~\bibnamefont{Botta}} \bibnamefont{et al.},
  \bibinfo{journal}{Phys. Rev. B} \textbf{\bibinfo{volume}{46}},
  \bibinfo{pages}{13008} (\bibinfo{year}{1992}).

\bibitem[{\citenamefont{Louarn et~al.}(1996)\citenamefont{Louarn, Trznadel, and
  Buisson}}]{Louarn_1996}
\bibinfo{author}{\bibfnamefont{G.}~\bibnamefont{Louarn}} \bibnamefont{et al.},
  \bibinfo{journal}{J. Phys. Chem.}
  \textbf{\bibinfo{volume}{100}}, \bibinfo{pages}{12532}
  (\bibinfo{year}{1996}).

\bibitem[{\citenamefont{Samoc}(2003)}]{Samoc_2003}
\bibinfo{author}{\bibfnamefont{A.}~\bibnamefont{Samoc}},
  \bibinfo{journal}{J. Appl. Phys.} \textbf{\bibinfo{volume}{94}},
  \bibinfo{pages}{6167} (\bibinfo{year}{2003}).

\bibitem[{\citenamefont{Rumbles et~al.}(1996)\citenamefont{Rumbles, Samuel, and
  Magnani}}]{Rumbles_1996}
\bibinfo{author}{\bibfnamefont{G.}~\bibnamefont{Rumbles}}
  \bibnamefont{et al.}, \bibinfo{journal}{Synth. Met.}
  \textbf{\bibinfo{volume}{76}}, \bibinfo{pages}{47} (\bibinfo{year}{1996}).

\bibitem[{\citenamefont{Yue, Berry, and
  McCullough}(1996)}]{Yue_1996}
\bibinfo{author}{\bibfnamefont{S.}~\bibnamefont{Yue}},
\bibinfo{author}{\bibfnamefont{G.~C.}~\bibnamefont{Berry}}~\bibnamefont{and}
\bibinfo{author}{\bibfnamefont{R.~D.}~\bibnamefont{McCullough}},
  \bibinfo{journal}{Macromolecules} \textbf{\bibinfo{volume}{29}},
  \bibinfo{pages}{933} (\bibinfo{year}{1996}).

\bibitem[{\citenamefont{Theander et~al.}(2001)\citenamefont{Theander, Svensson,
  and Ruseckas}}]{Theander_2001}
\bibinfo{author}{\bibfnamefont{M.}~\bibnamefont{Theander}} \bibnamefont{et al.},
  \bibinfo{journal}{Chem. Phys. Lett.} \textbf{\bibinfo{volume}{337}},
  \bibinfo{pages}{277} (\bibinfo{year}{2001}).

\bibitem[{\citenamefont{Cornil et~al.}(1998)\citenamefont{Cornil, dos Santos,
  and Crispin}}]{Cornil_1998}
\bibinfo{author}{\bibfnamefont{J.}~\bibnamefont{Cornil}}
  \bibnamefont{et al.}, \bibinfo{journal}{J. Am. Chem.
  Soc.} \textbf{\bibinfo{volume}{120}}, \bibinfo{pages}{1289}
  (\bibinfo{year}{1998}).

\bibitem[{\citenamefont{Manas and Spano}(1998)}]{Manas_1998}
\bibinfo{author}{\bibfnamefont{E.~S.} \bibnamefont{Manas}} \bibnamefont{and}
  \bibinfo{author}{\bibfnamefont{F.~C.} \bibnamefont{Spano}},
  \bibinfo{journal}{J. Chem. Phys.}
  \textbf{\bibinfo{volume}{109}}, \bibinfo{pages}{8087} (\bibinfo{year}{1998}).

\bibitem[{\citenamefont{Beljonne et~al.}(2000)\citenamefont{Beljonne, Cornil,
  and Silbey}}]{Beljonne_2000}
\bibinfo{author}{\bibfnamefont{D.}~\bibnamefont{Beljonne}} \bibnamefont{et al.},
  \bibinfo{journal}{J. Chem. Phys.}
  \textbf{\bibinfo{volume}{112}}, \bibinfo{pages}{4749} (\bibinfo{year}{2000}).

\bibitem[{\citenamefont{Westenhoff et~al.}(2006)\citenamefont{Westenhoff,
  Abrusci, and Feast}}]{Westenhoff_2006_AdvMater}
\bibinfo{author}{\bibfnamefont{S.}~\bibnamefont{Westenhoff}} \bibnamefont{et al.},
  \bibinfo{journal}{Adv. Mater.} \textbf{\bibinfo{volume}{18}},
  \bibinfo{pages}{1281} (\bibinfo{year}{2006}).

\bibitem[{\citenamefont{Spano}(2006)}]{Spano_2006_P3HT}
\bibinfo{author}{\bibfnamefont{F.~C.} \bibnamefont{Spano}},
  \bibinfo{journal}{Chem. Phys.} \textbf{\bibinfo{volume}{325}},
  \bibinfo{pages}{22} (\bibinfo{year}{2006}).

\bibitem[{\citenamefont{Chang et~al.}(2006)\citenamefont{Chang, Clark, and
  Zhao}}]{Chang_2006}
\bibinfo{author}{\bibfnamefont{J.~F.} \bibnamefont{Chang}} \bibnamefont{et al.},
  \bibinfo{journal}{Phys. Rev. B} \textbf{\bibinfo{volume}{74}},
  \bibinfo{pages}{115318}
  (\bibinfo{year}{2006}).
  
\end{thebibliography}
\end{document}